\documentclass[aps,twocolumn,nofootinbib,superscriptaddress,tightenlines]{revtex4}

\usepackage{epsfig,latexsym,cancel,amssymb,amsmath,verbatim,mathrsfs}
\usepackage{color,xcolor}
\usepackage{graphicx}
\usepackage{hyperref}
\hypersetup{
	colorlinks,
	linkcolor={red!70!black},
	citecolor={blue!80!black},
	urlcolor={blue!80!black},
	bookmarksopen=true
}

\newcommand{\nc}{\newcommand}
\nc{\beq}{\begin{equation}}  \nc{\eeq}{\end{equation}}
\nc{\bea}{\begin{eqnarray}}  \nc{\eea}{\end{eqnarray}}
\nc{\bel}{\begin{align}}         \nc{\eel}{\end{align}}
\nc{\nn}{\nonumber}

\begin{document}

\title{Single Top Quark Production with and without a Higgs Boson}

\author{Qing-Hong Cao}
\email{qinghongcao@pku.edu.cn}
\affiliation{School of Physics and State Key Laboratory 
of Nuclear Physics and Technology, Peking University, Beijing 100871, China}
\affiliation{Collaborative Innovation Center of Quantum Matter, Beijing 100871, China}
\affiliation{Center for High Energy Physics, Peking University, Beijing 100871, China}
\author{Hao-ran Jiang}
\email{H.R.Jiang@pku.edu.cn}
\affiliation{School of Physics and State Key Laboratory 
of Nuclear Physics and Technology, Peking University, Beijing 100871, China}
\author{Guojin Zeng}
\email{guojintseng@pku.edu.cn}
\affiliation{School of Physics and State Key Laboratory 
of Nuclear Physics and Technology, Peking University, Beijing 100871, China}

\begin{abstract}
One way to probe new physics beyond standard model is to check the correlation among higher dimension operators in effective field theory. We examine the strong correlation between the processes of $pp\rightarrow tHq$ and $pp\rightarrow tq$ which both depend on the same three operators. The correlation indicates that, according to the data of $pp\rightarrow tq$, $\sigma_{tHq}=\big[106.8 \pm 64.8\big]~{\rm fb}$ which is far below the current upper limit $\sigma_{tHq}\leq 900~{\rm fb}$.
\end{abstract}

\maketitle

\noindent{\bf 1. Introduction.}

\begin{figure}
  \centering
  \includegraphics[scale=0.315]{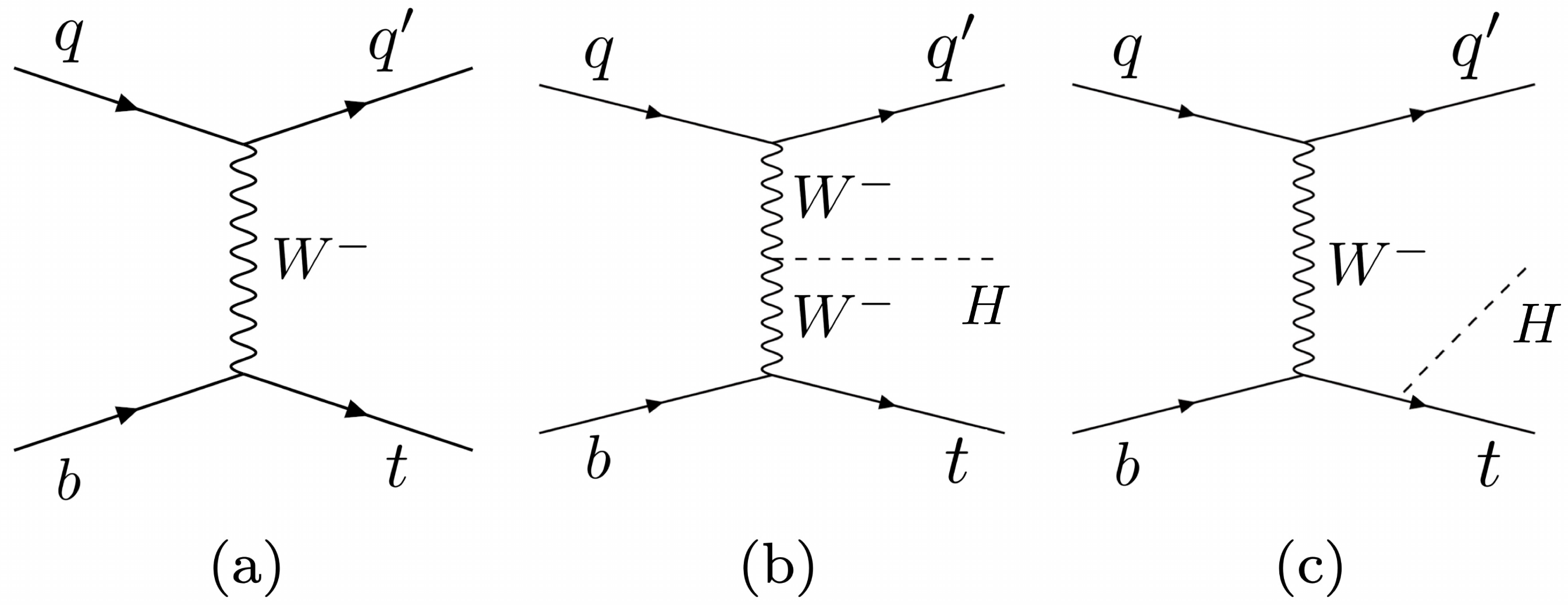}
  \caption{Feynman diagrams of the single-$t$ channel (a) and the $tHq$ production (b,~c).}
  \label{thq_singletop_process}
\end{figure}

One way to probe new physics (NP) beyond the Standard Model (SM) is from the so-called bottom-up approach, i.e. one describes the unknown NP effects through high-dimensional operators constructed with the SM fields at the NP scale $\Lambda$, obeying the well-established gauge structure of the SM, i.e. $SU(2)_{W}\otimes U(1)_Y$. 
The Lagrangian of effective field theory (EFT) is
\begin{equation}
  \mathscr{L}_{\rm EFT}=\mathscr{L}_{\rm SM}+\frac{1}{\Lambda^2}\sum_i \mathcal{C}^i O_i^{(6)}
  +O\left(\frac{1}{\Lambda^4}\right),
\end{equation}
where $\mathcal{C}_i$'s are Wilson coefficients. The aim of EFT analysis is to find nontrivial correlations among $\mathcal{C}_i$'s as those relations would shed lights on the structure of NP models. Various basis of dimension-6 operators have been introduced in the literature, e.g. Warsaw basis~\cite{Grzadkowski:2010es}, HISZ basis~\cite{PhysRevD.48.2182,Hagiwara_1997}, and SILH basis~\cite{Giudice_2007,Mass__2013}. Unfortunately, each basis consists of 59 independent operators, and it is difficult to explore the correlations among 59 operators in practice. One way out of the predicament is to find independent observables that are sensitive to a small set of operators and then examine correlations among those operators. That has been studied in single top productions~\cite{Cao_2007, AguilarSaavedra:2008zc, AguilarSaavedra:2009mx, Zhang:2010dr,Degrande_2018,Cao:2015doa,Xie:2021xtl}. In this work we explore the strong correlation among the $t$-channel single top (single-$t$) production and the associated production of single top quark with a Higgs boson (named as the $tHq$ production); see Fig.~\ref{thq_singletop_process}.  
The two channels mainly involve three operators which can be measured in the single-$t$ production. That yields a strong constraint on the cross section of the $tHq$ production.

As some high-dimensional operators can be absorbed by field redefinition, various schemes of input parameters has been widely discussed in the literature~\cite{Brivio_2019,Brivio:2017btx}. We choose the Warsaw basis~\cite{Grzadkowski:2010es} and follow the scheme of input parameters proposed in~\cite{SMEFTFR:2017zog} to normalize the SM fields and masses. Several approximations made in our analysis are listed as follows: i) only diagonal elements of CKM matrix have been considered for they are much larger than off-diagonal elements; ii) we ignore the masses of light quarks  and keep only the top quark mass hereafter; iii) gauge couplings of light quarks are set to be the same as the SM as these couplings are strictly constrained by hadron experiments, iv) the influence of branch ratio from top quark decay is neglected since the effects on partial width and total width will cancel with each other, leaving a next leading order contribution; v) CP conservation is supposed to avoid complex Wilson coefficients.

The SMEFT analysis in the single-$t$ channel has been carried out in the literature~\cite{Cao_2007, AguilarSaavedra:2008zc, AguilarSaavedra:2009mx, Zhang:2010dr}. Only 4 independent dimension-6 operators are involved in the $t$-channel production: 
\begin{align}
  O_{uW} &= (\bar q_p \sigma^{\mu\nu} u_r) \tau^I \tilde\phi W^{I}_{\mu\nu} + h.c. ,\nn\\
  O_{\phi q3} &= (\phi^\dagger i \stackrel{\leftrightarrow}{D^I_\mu}\phi)(\bar q_p \tau^I \gamma^\mu q_r) + h.c. ,\nn\\
  O^{(1)}_{qq} &= (\bar q_p \gamma_\mu q_r) (\bar q_s \gamma^\mu q_t) + h.c. , \nn\\
  O^{(3)}_{qq} &= (\bar q_p \gamma_\mu \tau^I q_r)  (\bar q_p \gamma^\mu \tau^I q_r) + h.c.\, .
\end{align}
where $\phi$ denotes the SM Higgs boson doublet, $D_{\mu}$ the covariant
derivative, $q_i$ the left-handed $SU(2)$ doublet of the $i$-th generation,
and $u_r$ the right-handed isosinglets~\citep{Buchmuller:1985jz};
$\tau^{I}$ denote the usual Pauli matrices in the weak isospin space. 
Considering the flavor structure of four-fermion operators, there can be flavor changing currents from both the first and the second generations to the third generation. However, we would ignore the contribution from the second generation, which suffers a suppression from the PDF. With Fierz Identity, it can be proved that only two independent flavored four-fermion operators have contribution to the $t$-channel production. They are
\begin{align}
  O^{(1)}_{3113} &= (\bar q_3 \gamma_\mu q_1) (\bar q_1 \gamma^\mu q_3) + h.c., \nn\\
  O^{(3)}_{3311} &= (\bar q_3 \gamma_\mu \tau^I q_3)  (\bar q_1 \gamma^\mu \tau^I q_1) + h.c.\, .
\end{align}
For convenience, we will still use $O^{(1)}_{qq}$ and $O^{(3)}_{qq}$ to denotes these two flavored operators respectively. Indeed, the interference between $O^{(3)}_{qq}$ and the SM is only three times as large as that of $O^{(1)}_{qq}$ due to a color factor. Although their contributions can be separated by quadratic terms, they can be considered as a same degree of freedom in the leading order analysis, i.e. keeping only the interference between the SM and NP operators. We thus use $O^{(3)}_{qq}$ to denote the degree of  freedom hereafter.

The cross sections of the $t$-channel single top-quark productions at the 13 TeV LHC are  
\begin{align}
\label{eq:sigmat}
&\sigma_t 
= \left[214 -13C_{qq}^{(3)} + 16C_{uW} + 13C_{\phi q3} \right] {\rm pb},\nonumber\\
&\sigma_{\bar{t}} = \left[81 - 4 C_{qq}^{(3)} + 5 C_{uW} + 5 C_{\phi q3} \right] {\rm pb},
\end{align}
where 
\begin{align}
C_{qq}^{(3)}&\equiv \mathcal{C}_{qq}^{(3)}\left(\tfrac{\rm TeV}{\Lambda^2}\right)^2,\nn\\
C_{uW} &\equiv\mathcal{C}_{uW}\left(\tfrac{\rm TeV}{\Lambda^2}\right)^2,\nn\\
C_{\phi q3} &\equiv \mathcal{C}_{\phi q 3}\left(\tfrac{\rm TeV}{\Lambda^2}\right)^2,\nn
\end{align}
and the first constant term denotes the SM contribution with NNLO QCD corrections~\cite{Gao:2020ejr}.  
Both FeynRules and MadGraph packages are used in our calculation~\cite{Alwall:2014hca,Alloul:2013bka}.

\begin{figure}
  \centering
  \includegraphics[scale=0.2]{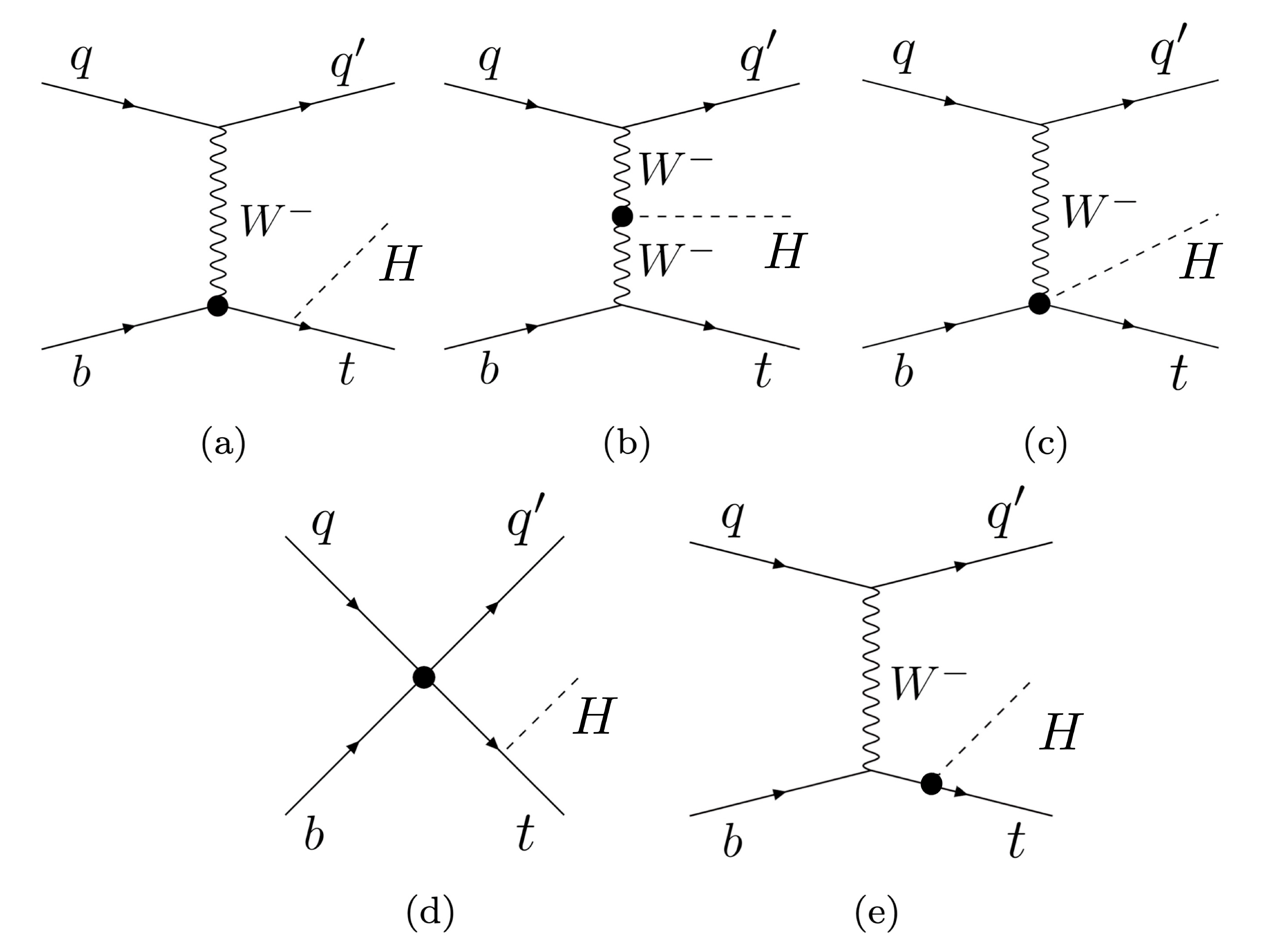}
  \caption{The $tHq$ production induced by effective operators}
  \label{thq_NP_diagrams}
\end{figure}

The $tHq$ production involves more operators  than the single-$t$ channel.  Figure~\ref{thq_NP_diagrams} shows possible modifications to the $tHq$ production from various operators; see the black thick dots. 
In addition to the operators affecting the single-$t$ production, the $tHq$ production involves operators as follows~\cite{Grzadkowski:2010es}:
\begin{align}
O_{\phi D} &= (\phi^\dagger D^\mu \phi)^*(\phi^\dagger D_\mu \phi),\nn\\
O_{u\phi} &= (\phi^\dagger \phi)(\bar q_p u_r \tilde\phi)+h.c.~,\nn\\
O_{\phi \Box} &= (\phi^\dagger \phi)\Box(\phi^\dagger \phi),\nn\\
O_{\phi W} &= \phi^\dagger \phi W^I_{\mu\nu} W^{I\mu\nu}.
\end{align}
The operator $O_{u\phi}$ is tightly constrained from gluon-gluon fusion measurements~\cite{Dordevic:2019fft}, and the operator $O_{\phi W}$ is severely bounded by the combined analysis of the $H \gamma \gamma$, $HZZ$ and $HZ \gamma$ couplings~\cite{Dawson:2013bba}. The operators $O_{\phi D}$ and $O_{\phi \Box}$ are  constrained by the oblique parameters~\cite{Peskin:1991sw,Han:2004az} and also are mildly sensitive to the $tHq$ production~\cite{Degrande_2018}. 
Therefore, when keeping only the interference effect between the SM and NP channels, we end up with only three independent operators, $O^{(3)}_{qq}$ , $O_{\phi q3}$ and $O_{uW}$, which affect both the single-$t$ production and the $tHq$ production.  

The $tHq$ production is not found yet owing to its small production rate. We sum up both the $tHq$ and $\bar{t}Hq^\prime$ productions hereafter and denote the production cross section of both the $tHq$ and $\bar{t}Hq^\prime$ productions as $\sigma_{tHq}$, which is
\begin{align}
\sigma_{tHq} =\left[74.3-11.3C_{qq}^{(3)} + 22.0C_{uW} - 2.6 C_{\phi q3}\right] {\rm fb}.
\end{align}

\noindent{\bf 2. Collider phenomenology.}

Thanks to the large production rate of the single-$t$ production, bot the total cross section and the differential distributions are well measured at the LHC and can be used to search for new physics beyond the SM.  The leptonic decays of the top quark provide clean collider signature at the LHC and are used widely in experimental searches. We thus focus on the channel of $t\rightarrow \mu + \nu_\mu + b$ and choose ${\rm Br}(t\rightarrow b\mu  \nu_\mu ) =(13.4\pm0.6)\%$~\cite{Tanabashi:2018oca}. 

Three observables are needed to probe the  three operators $O^{(3)}_{qq}$, $O_{\phi q3}$ and $O_{uW}$. In our analysis we consider both the single top-quark production and the single antitop-quark production. The first observable in our analysis is the total cross section of the $t$-channel single-top production, i.e. 
\beq
\sigma_{t+\bar{t}}=\sigma(tq)+\sigma(\bar{t}q).
\eeq
The second observable is the cross section ratio of the single-$t$ production and single-$\bar{t}$ production, defined as
\begin{equation}
  R_{t} \equiv \frac{\sigma_{t}}{\sigma_{\bar{t}}}.
\end{equation}
The difference in production rates mainly arises from the parton distribution functions (PDFs) of initial state. 

The third observable is based on the spin correlation between the charged lepton and top-quark. Owing to its large production rate  the single top-quark processes are well measured at the LHC such that one can examine differential distributions of various observables. We adopt the so-called ``spectator basis" to maximize spin correlations by taking advantage of the fact that the top quark produced through the single-$t$ processes is almost $100\%$ polarized along the direction of the spectator quark, the light jet produced in association with the top quark~\cite{Kane:1991bg,Mahlon:1996pn,Cao:2004ky,Sullivan:2004ie,Cao:2005pq}. The distribution of the cosine of the angle between the charged lepton and the spectator jet in the rest frame of top quark is 
\beq
  \frac{1}{\sigma} \frac{\mathrm{d} \sigma}{\mathrm{d} \cos\theta} = \frac{1}{2}\left(1+\frac{N_- -N+}{N_-+N_+} \cos\theta \right),
\eeq
where $N_-$ ($N_+$) is the number of top quark with polarization against (along) the direction of the spectator jet, respectively. The third observable $A_{FB}$ is defined as
\begin{equation}
A_{FB}=\frac{\sigma_F-\sigma_B}{\sigma_F+\sigma_B} 
\end{equation}
where 
\begin{equation}
\sigma_F\equiv \int_0^1 \frac{\mathrm{d} \sigma}{\mathrm{d} \cos\theta} \mathrm{d} \cos\theta,~~\sigma_B\equiv \int_{-1}^0 \frac{\mathrm{d} \sigma}{\mathrm{d} \cos\theta} \mathrm{d} \cos\theta.
\end{equation}
The distribution of $\cos\theta$ is modified by the operators as following:
\beq
\begin{aligned}
 \frac{\mathrm{d} \sigma}{\mathrm{d} \cos\theta} & = \frac{1}{2}\left(\sigma_{t}^{0} + \sum_{i} C_i \sigma_{t}^{i} \right)\\
 &+ \left( A^{0}_{FB} \sigma_{t}^{0} + \sum_{i} C_i A^i_{FB} \sigma^i_t \right) \cos\theta,
\end{aligned}
\eeq
which yields 
\beq
A_{FB}  = \frac{A^{0}_{FB} \sigma_{t}^{0} + \sum_{i=1}^{3} C_i A^i_{FB} \sigma^i_t}{\sigma_{t}^{0} + \sum_{i=1}^{3} C_i \sigma_{t}^{i}}.
\eeq
where $\sigma_t^{i}$ and $A_{FB}^{i}$ represent the cross section and asymmetry generated by the SM ($i=0$) and heavy operators ($i=1,2,3)$. We take both the single-$t$ and single-$\bar{t}$ into account in the $\sigma_t$ and $A_{FB}$ analysis. The numerical values of the $\sigma_t^{i}$'s can be read out from Eq.~\ref{eq:sigmat}, and the values of $A_{FB}$'s are listed as follows:
\begin{align}\label{AFB}
&A_{FB}^0 = 0.4596, && A_{FB}^{O_{\phi q3}}=0.4596,\nonumber\\
&A_{FB}^{O_{qq}^{(3)}}= 0.4721, && A_{FB}^{O_{uW}}=0.3548,
\end{align}
The operator $O_{\phi q3}$ modifies only the magnitude of the $g_W$ coupling and do not affect the shape of all the differential distributions, therefore, it yields exactly the same asymmetry $A_{FB}$ as the SM.

In our analysis we normalize the three observables by the SM predictions as follows:
\beq
  \bar{\sigma}_{t}  \equiv \frac{\sigma_{t+\bar{t}}}{\sigma_{t+\bar{t}}^0}, ~~ \mathcal{A}  \equiv \frac{A_{FB}}{A_{FB}^0}, ~~ \mathcal{R}  \equiv \frac{R_{t}}{R_{t}^0}.
\eeq
Since the contribution of NP operators are assumed to be smaller than the SM, keeping only the linear term of Wilson coefficients yields
\begin{eqnarray}
&& \bar{\sigma}_{t} = 1 - 0.059 \times C_{qq}^{(3)} + 0.075\times C_{uW}+ 0.060\times C_{\phi q3},\nonumber\\
&&{\mathcal{A}} = 1 - 0.0033 \times C_{qq}^{(3)} - 0.0349 \times C_{uW},\nonumber\\
&& \mathcal{R} = 1 - 0.0308 \times C_{qq}^{(3)} + 0.0328 \times C_{uW},
\end{eqnarray}
from which we obtain 
\begin{align}
  \label{wilson_observable}
    C_{qq}^{(3)}  &= 57 - 28\mathcal{A} - 29 \mathcal{R}, \nn\\
    C_{uW}  &= 23 - 26 \mathcal{A} + 3\mathcal{R}, \nn\\
    C_{\phi q3} & = 10+ 17 \bar\sigma_{t} + 6 \mathcal{A} - 33\mathcal{R}. 
  \end{align}
Note that the three observables exhibit quite distinct sensitivity to the three operators when considering only the leading effect of NP operators. First, the operator $O_{\phi q 3}$ only modify the magnitude of the $W$-$t$-$b$ coupling and does not alter any differential distribution, therefore, it does not contribute to either the asymmetry $\mathcal{A}$ or the charge ratio $\mathcal{R}$. Second, the asymmetry $\mathcal{A}$ depends mainly on $O_{uW}$ owing to the fact that $O_{uW}$ flips the chirality of the top quark and yields an obvious deviation in the $\cos\theta$ distribution. Third, both the $O_{qq}^{(3)}$ and $O_{uW}$ yields comparable interference contribution to the single-$t$ production and thus contribute equally to the $\mathcal{R}$ which is more sensitive to the parton distributions of valence and sea quarks in the initial state. Finally, the cross section $\bar{\sigma}_t$ is sensitive to all the three operators.

Combined analysis of the three observables help us probing NP beyond the SM. For example, the relation of $C_{uW}$ and $C^{(3)}_{qq}$ can be determined from the $\mathcal{A}$ and $\mathcal{R}$ measurements; see Eq.~\ref{wilson_observable}. In particular, if $\mathcal{R}\simeq 1$ and $\mathcal{A}\neq 1$, then $C_{uW}\simeq C_{qq}^{(3)}\neq 0$; if $\mathcal{A}\simeq 1$ and $\mathcal{R}\neq 1$, then $C_{qq}^{(3)}\sim -10C_{uW}$. If no deviations are found in the $\mathcal{A}$ and $\mathcal{R}$ measurements, say $\mathcal{A}\simeq \mathcal{R}\simeq 1$, then the only possible NP source in the single-$t$ process will be the $O_{\phi q3}$ which could be obtained from the $\bar{\sigma}_{t}$.

The current bound of the total cross section of the single-$t$ production is $176~\text{pb}\leq \sigma_t\leq 238~\text{pb}$ at the $1\sigma$ confidence level~\cite{Sirunyan:2018rlu}, and the relative error of the cross section measurement is $\delta \sigma_t=15\%$. It yields 
\begin{equation}
\bar{\sigma}_{t}=0.95\pm0.14~.
\end{equation}
The charge ratio~\cite{Aaboud:2016ymp, Sirunyan:2018rlu, CMS:2019dvz}  and the $\cos\theta$ distribution~\cite{ CMS:2019dvz, Khachatryan:2015dzz, Tiko:2016brs}  in the single top processes have been measured at the 13~TeV LHC. Taking advantage of the unfolding results on the charge ratio and angular distribution provided by the CMS collaboration~\cite{Khachatryan:2015dzz,CMS:2019dvz}, we compare our parton level analysis directly to the experimental data and obtain
\begin{align} 
\mathcal{R}&=1\pm 0.035,\\
\mathcal{A} &= 0.877\pm 0.154,
\label{eq:R}
\end{align}
where the $\mathcal{A}$ is obtained from the linear fitting of the $A_\mu$ distribution in Ref.~\cite{CMS:2019dvz}.
That gives rise to the constraints on the three operators as follows:
\begin{align}
& -1.9\leq C_{qq}^{(3)} \leq 8.8~,\nn\\
& -0.9\leq C_{uW} \leq 7.3~,\nn\\
& -5.2\leq C_{\phi q3} \leq 3.9~.
\end{align}

\begin{figure*}
\includegraphics[scale=0.275]{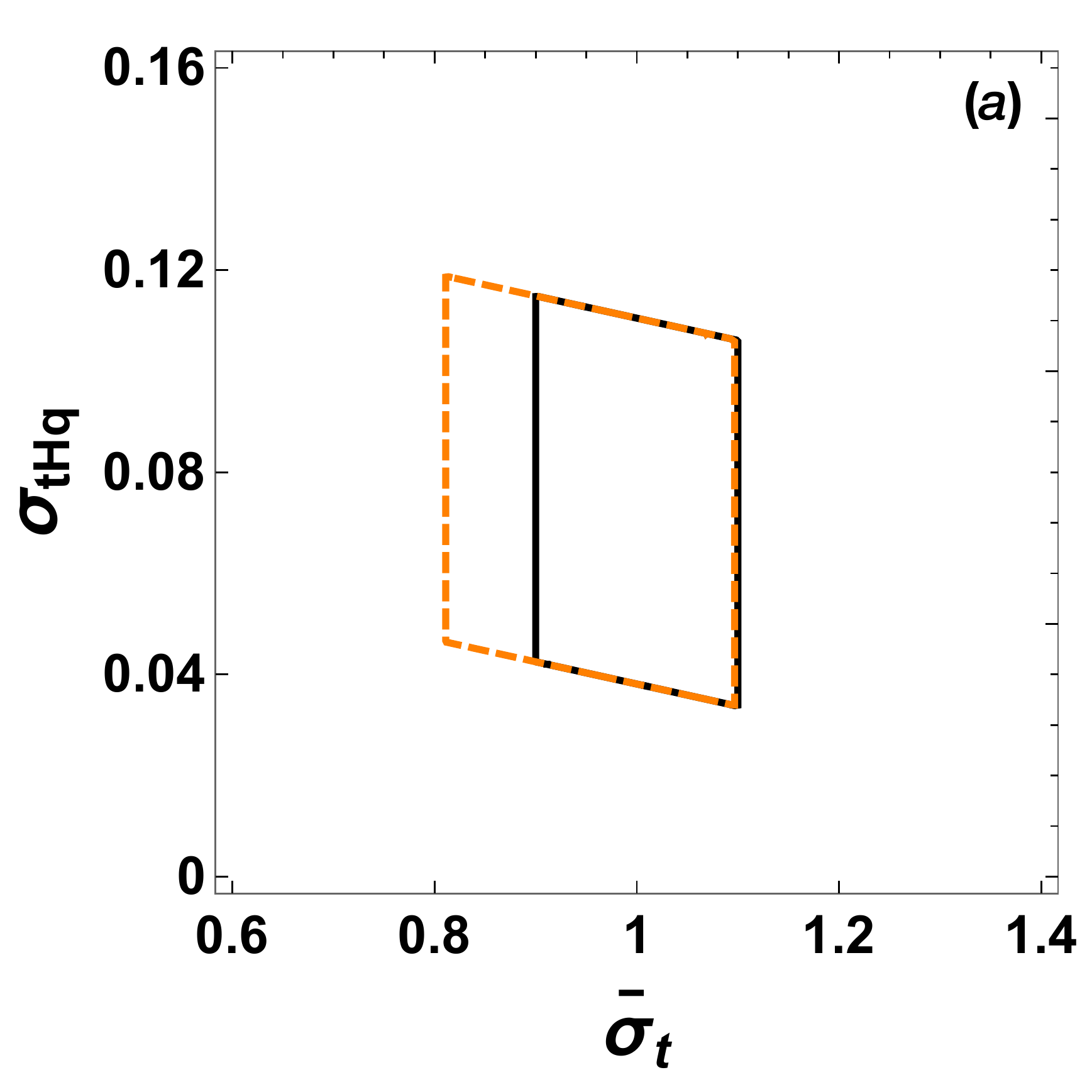}
\includegraphics[scale=0.275]{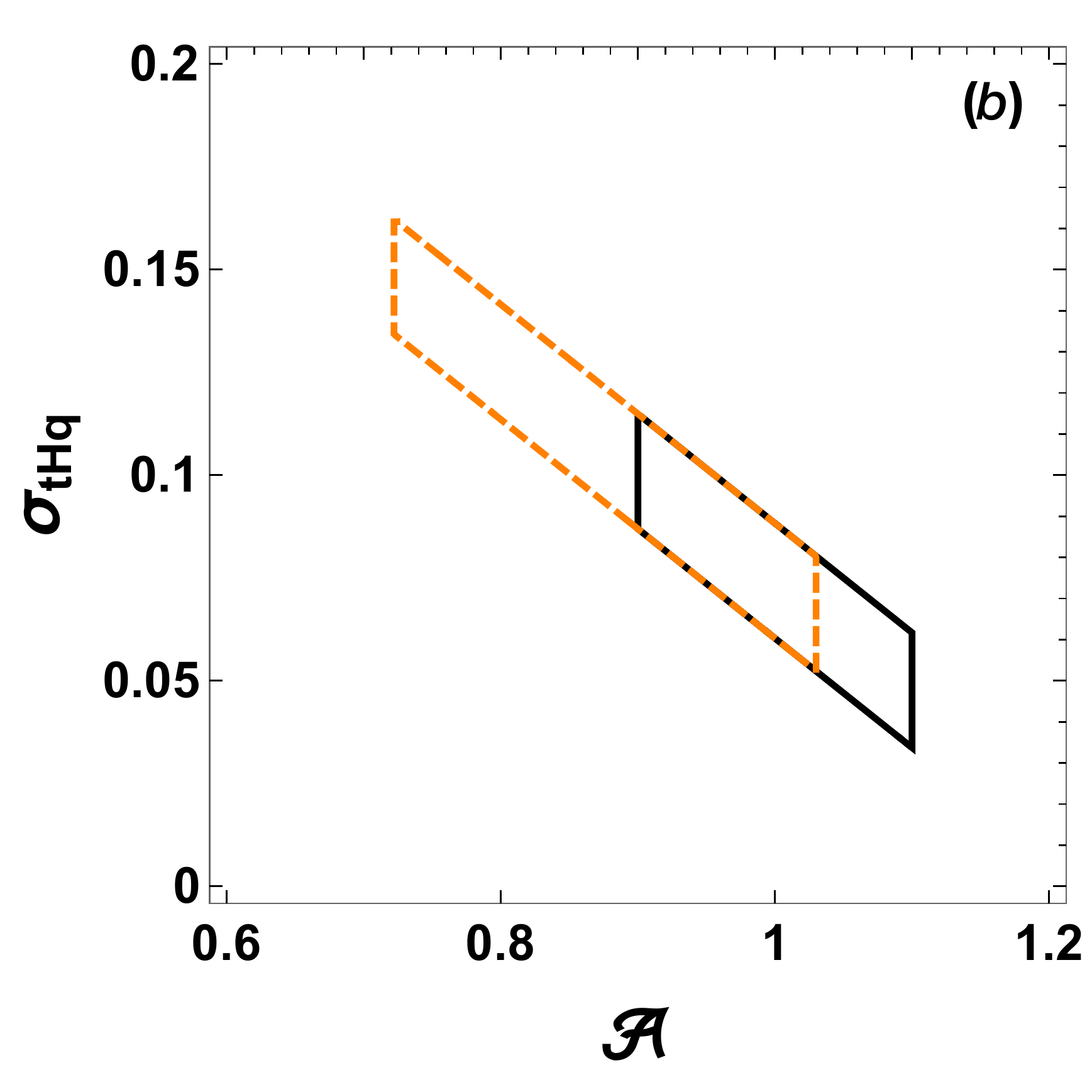}
\includegraphics[scale=0.275]{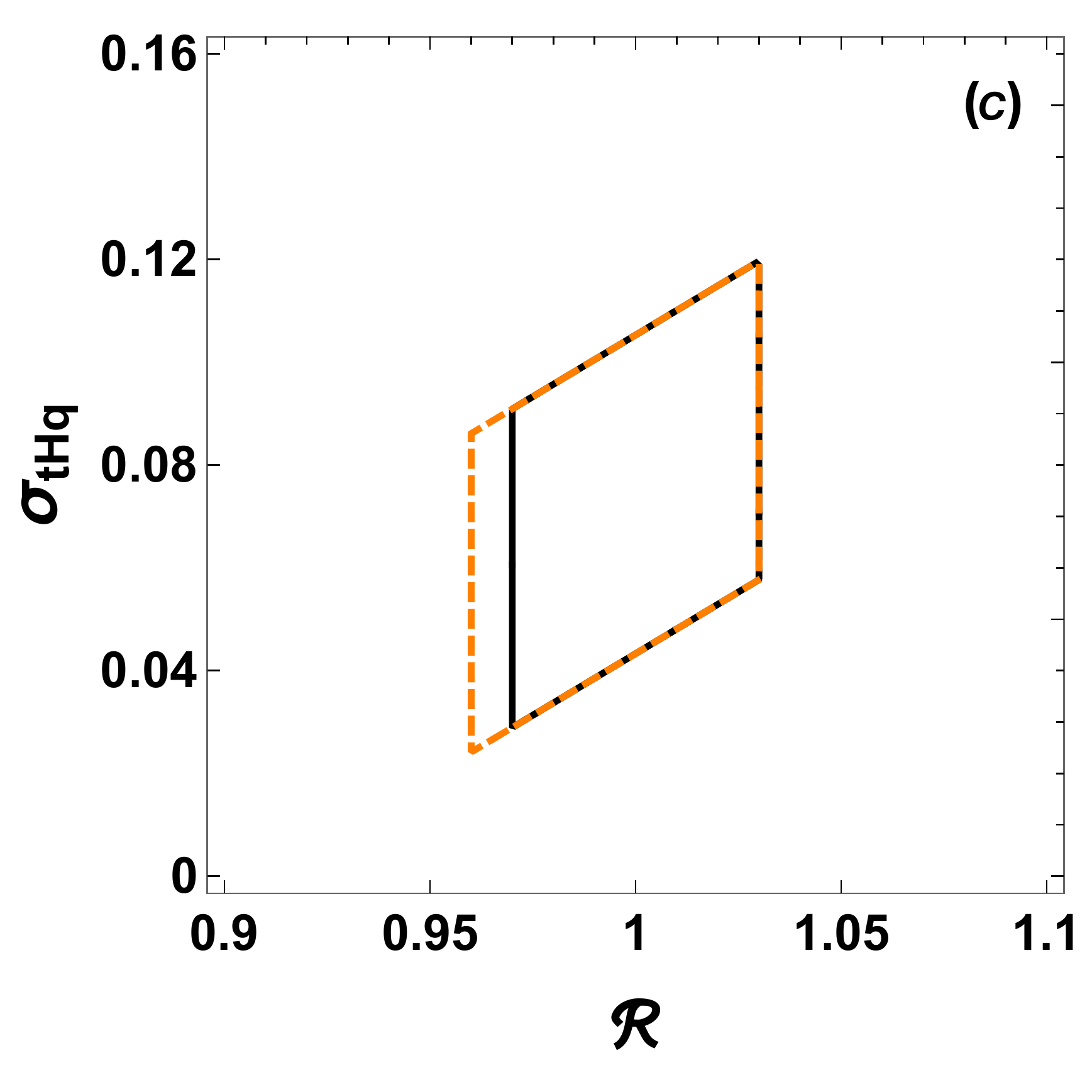}
 \caption{Correlations among $\sigma_{tHq}$ and the cross section $\bar{\sigma}_t$ (a) , the asymmetry $\mathcal{A}$ (b), and the charge ratio $\mathcal{R}$ (c) at the $1\sigma$ confidence level at the current 13~TeV LHC  (orange) and the HL-LHC (black).  The central values of all the three observables are chosen as 1 at the HL-LHC.}
\label{sigma_thq_observable}
\end{figure*}

Even though the current measurements impose only rough bounds on operators, more accurate measurements are anticipated at the high luminosity LHC with an integrated luminosity of $3000~{\rm fb}^{-1}$ (HL-LHC). The uncertainties are mainly due to the parton distribution functions and the systematic errors. The major uncertainties of the cross section and forward-backward asymmetry measurements are due to the systematic error which are roughly about 15\%~\cite{Sirunyan:2018rlu}. On the other hand, the statistical uncertainty is about 1\% which can be ignored in our analysis.  Therefore, we expect the uncertainties of the cross section and $A_{FB}$ measurements to be reduced to 10\% in our projection of the HL-LHC potential of measuring the single-$t$ channel.

In the charge ratio measurement the large systematic errors tend to cancel out and give rise to a net value comparable to the PDF uncertainty, e.g. both the PDF and systematic errors are $\sim 3.0\%$ in the recent CMS data~\cite{Sirunyan:2018rlu}. In accord to the NNPDF study the PDF error is expected to less than 3\% when more precision measurements are available~\cite{Nocera:2019wyk}. Therefore, we optimistically expect the charge ratio to be measured with an accuracy of 3\% in the forthcoming HL-LHC.

Assuming the central values of the three operators are the same as the SM predictions, i.e. centralizing around zero, we obtain the projected constraints on the operators at the HL-LHC as follows:  
\begin{align}
& -3.7\leq C_{qq}^{(3)} \leq 3.7~,\nn\\
& -2.7\leq C_{uW} \leq 2.7~,\nn\\
& -3.2\leq C_{\phi q3} \leq 3.2~,
\end{align}
where both the cross section and $A_{FB}$ measurement are of 10\% uncertainties, 
and more optimistically, 
\begin{align}
& -2.3\leq C_{qq}^{(3)} \leq 2.3~,\nn\\
& -1.4\leq C_{uW} \leq 1.4~,\nn\\
& -2.1\leq C_{\phi q3} \leq 2.1~,
\end{align}
when the systematic uncertainties are improved to be 5\%.

\noindent{\bf 4. Single-$t$ channel versus $tHq$ channel}

We now examine the correlation between the single-$t$ production and the $tHq$ production as both the single-$t$ channel and the $tHq$ channel mainly depend on the three operators. A simple algebra gives rise to 
\begin{equation}
\label{thq_observable}
\sigma_{tHq} = \Big[-95.1 - 44.0\bar{\sigma}_{t}- 266.0{\mathcal{A}} +479.4\mathcal{R}\Big]~ \text{fb}, 
\end{equation}
which serves well for checking the consistence of the experimental measurement and operator analysis. It tells us that the information of the $tHq$ production rate can be inferred from the single-$t$ production as long as one keeps the interference effect and neglects those sub-leading operators explained above.  For example, based on the current measurement of the single-$t$ production, our operator analysis shows that 
\begin{equation}
\sigma_{tHq}=\Big[106.8 \pm 64.8\Big]~{\rm fb}
\end{equation}
at the $1\sigma$ confidence level, which is far below the current upper limit $\sigma_{tHq}\leq 900~{\rm fb}$~\cite{Sirunyan:2018lzm, Aad:2020ivc}.

Figure~\ref{sigma_thq_observable} displays the correlations among $\sigma_{tHq}$ and the cross section $\bar{\sigma}_t$ (a) , the asymmetry $\mathcal{A}$ (b), and the charge ratio $\mathcal{R}$ (c) at the $1\sigma$ confidence level at the current 13~TeV LHC  (orange) and the HL-LHC (black).  In the study of the HL-LHC, the central values of all the three observables are chosen as 1, the uncertainties of the $\sigma_{t}$ and $A_{FB}$ are set to be 10\%, and the uncertainty of the charge ratio $\mathcal{R}$ is 3\%. The capability of the operator analysis is highly limited by the large systematic uncertainties in the cross section $\bar{\sigma}_t$ and the asymmetry $\mathcal{A}$. The strongest bound on $\sigma_{tHq}$ is derived from the charge ratio $\mathcal{R}$ measurement  owing to the small systematic error.  

Assuming the central value of $\sigma_{tHq}$ is still the same as the SM prediction at the 13~TeV HL-LHC, we obtain the projected $\sigma_{tHq}$ as following: 
\begin{align}
\sigma_{tHq}&= \Big[74.3 \pm 45.4~(29.9)\Big]~{\rm fb},
\end{align}
where the systematic errors of both the cross section and $A_{FB}$ measurements are $10\%$ (5\%), respectively,
and the error of charge ratio measurement is 3\%.

\noindent{\bf 5. Conclusion.}

Effective field theory is a very powerful tool to probe new physics beyond the Standard Model in a model-independent approach. More accurate information of higher dimension operators are anticipated at the large hadron collider in the operation phase of high luminosity. When new physics resonances decouple from the electroweak scale, they leave their footprints in various relations among higher-dimension operators, i.e., strong correlations among the Wilson coefficients. We pointed out that the single-$t$ production and the $tHq$ channel are highly correlated as both processes depend mainly on three dimension-6 operators. At the leading order of operator expansion, we obtain a relation as following: 
\begin{align}
\sigma_{tHq} = \bigg[ &-95.1 - 44.0\times \frac{\sigma_{t+\bar{t}}}{\sigma^{\rm SM}_{t+\bar{t}}}\nn\\
&- 266.0\times \frac{A_{\rm FB}}{A_{\rm FB}^{\rm SM}} +479.4\times \frac{R_t}{R_t^{\rm SM}}\bigg]~ \text{fb}, \nn
\end{align}
where $\sigma_{t+\bar{t}}$, $A_{\rm FB}$ and $R_t=\sigma_{t}/\sigma_{\bar t}$ denotes the total cross section, the asymmetry and the charge ratio measured in the single-$t$ processes, respectively. 
The relation predicts that, according to the current data of single-$t$ channel, the yet-to-be measured cross section of the $tHq$ production at the 13~TeV LHC is 
\begin{equation}
\sigma_{tHq}=\Big[106.8 \pm 64.8\Big]~{\rm fb}
\end{equation}
at the $1\sigma$ confidence level, which is far below the current upper limit $\sigma_{tHq}\leq 900~{\rm fb}$.
Assuming the central value of $\sigma_{tHq}$ is still the same as the SM prediction at the 13~TeV HL-LHC, we obtain the projected $\sigma_{tHq}$ as following: 
\begin{align}
\sigma_{tHq}&= \Big[74.3 \pm 45.4\Big]~{\rm fb},
\end{align}
where the systematic errors of both the cross section and $A_{FB}$ measurements are $10\%$ and the error of charge ratio measurement is 3\%.

\noindent{\bf Acknowledgments.}
We thank Yandong Liu and Rui Zhang for enlightening discussions and comments. The work is supported in part by the National Science Foundation of China under Grant Nos. 11725520, 11675002, and 11635001.

\bibliographystyle{apsrev}
\bibliography{reference}

\begin{thebibliography}{37}
\expandafter\ifx\csname natexlab\endcsname\relax\def\natexlab#1{#1}\fi
\expandafter\ifx\csname bibnamefont\endcsname\relax
  \def\bibnamefont#1{#1}\fi
\expandafter\ifx\csname bibfnamefont\endcsname\relax
  \def\bibfnamefont#1{#1}\fi
\expandafter\ifx\csname citenamefont\endcsname\relax
  \def\citenamefont#1{#1}\fi
\expandafter\ifx\csname url\endcsname\relax
  \def\url#1{\texttt{#1}}\fi
\expandafter\ifx\csname urlprefix\endcsname\relax\def\urlprefix{URL }\fi
\providecommand{\bibinfo}[2]{#2}
\providecommand{\eprint}[2][]{\url{#2}}

\bibitem[{\citenamefont{Grzadkowski et~al.}(2010)\citenamefont{Grzadkowski,
  Iskrzynski, Misiak, and Rosiek}}]{Grzadkowski:2010es}
\bibinfo{author}{\bibfnamefont{B.}~\bibnamefont{Grzadkowski}},
  \bibinfo{author}{\bibfnamefont{M.}~\bibnamefont{Iskrzynski}},
  \bibinfo{author}{\bibfnamefont{M.}~\bibnamefont{Misiak}}, \bibnamefont{and}
  \bibinfo{author}{\bibfnamefont{J.}~\bibnamefont{Rosiek}},
  \bibinfo{journal}{JHEP} \textbf{\bibinfo{volume}{10}}, \bibinfo{pages}{085}
  (\bibinfo{year}{2010}), \eprint{1008.4884}.

\bibitem[{\citenamefont{Hagiwara et~al.}(1993)\citenamefont{Hagiwara, Ishihara,
  Szalapski, and Zeppenfeld}}]{PhysRevD.48.2182}
\bibinfo{author}{\bibfnamefont{K.}~\bibnamefont{Hagiwara}},
  \bibinfo{author}{\bibfnamefont{S.}~\bibnamefont{Ishihara}},
  \bibinfo{author}{\bibfnamefont{R.}~\bibnamefont{Szalapski}},
  \bibnamefont{and}
  \bibinfo{author}{\bibfnamefont{D.}~\bibnamefont{Zeppenfeld}},
  \bibinfo{journal}{Phys. Rev. D} \textbf{\bibinfo{volume}{48}},
  \bibinfo{pages}{2182} (\bibinfo{year}{1993}),
  \urlprefix\url{https://link.aps.org/doi/10.1103/PhysRevD.48.2182}.

\bibitem[{\citenamefont{Hagiwara et~al.}(1997)\citenamefont{Hagiwara,
  Hatsukano, Ishihara, and Szalapski}}]{Hagiwara_1997}
\bibinfo{author}{\bibfnamefont{K.}~\bibnamefont{Hagiwara}},
  \bibinfo{author}{\bibfnamefont{T.}~\bibnamefont{Hatsukano}},
  \bibinfo{author}{\bibfnamefont{S.}~\bibnamefont{Ishihara}}, \bibnamefont{and}
  \bibinfo{author}{\bibfnamefont{R.}~\bibnamefont{Szalapski}},
  \bibinfo{journal}{Nuclear Physics B} \textbf{\bibinfo{volume}{496}},
  \bibinfo{pages}{66–102} (\bibinfo{year}{1997}), ISSN
  \bibinfo{issn}{0550-3213},
  \urlprefix\url{http://dx.doi.org/10.1016/S0550-3213(97)00208-3}.

\bibitem[{\citenamefont{Giudice et~al.}(2007)\citenamefont{Giudice, Grojean,
  Pomarol, and Rattazzi}}]{Giudice_2007}
\bibinfo{author}{\bibfnamefont{G.~F.} \bibnamefont{Giudice}},
  \bibinfo{author}{\bibfnamefont{C.}~\bibnamefont{Grojean}},
  \bibinfo{author}{\bibfnamefont{A.}~\bibnamefont{Pomarol}}, \bibnamefont{and}
  \bibinfo{author}{\bibfnamefont{R.}~\bibnamefont{Rattazzi}},
  \bibinfo{journal}{Journal of High Energy Physics}
  \textbf{\bibinfo{volume}{2007}}, \bibinfo{pages}{045–045}
  (\bibinfo{year}{2007}), ISSN \bibinfo{issn}{1029-8479},
  \urlprefix\url{http://dx.doi.org/10.1088/1126-6708/2007/06/045}.

\bibitem[{\citenamefont{Massó and Sanz}(2013)}]{Mass__2013}
\bibinfo{author}{\bibfnamefont{E.}~\bibnamefont{Massó}} \bibnamefont{and}
  \bibinfo{author}{\bibfnamefont{V.}~\bibnamefont{Sanz}},
  \bibinfo{journal}{Physical Review D} \textbf{\bibinfo{volume}{87}}
  (\bibinfo{year}{2013}), ISSN \bibinfo{issn}{1550-2368},
  \urlprefix\url{http://dx.doi.org/10.1103/PhysRevD.87.033001}.

\bibitem[{\citenamefont{Cao et~al.}(2007)\citenamefont{Cao, Wudka, and
  Yuan}}]{Cao_2007}
\bibinfo{author}{\bibfnamefont{Q.-H.} \bibnamefont{Cao}},
  \bibinfo{author}{\bibfnamefont{J.}~\bibnamefont{Wudka}}, \bibnamefont{and}
  \bibinfo{author}{\bibfnamefont{C.-P.} \bibnamefont{Yuan}},
  \bibinfo{journal}{Physics Letters B} \textbf{\bibinfo{volume}{658}},
  \bibinfo{pages}{50–56} (\bibinfo{year}{2007}), ISSN
  \bibinfo{issn}{0370-2693},
  \urlprefix\url{http://dx.doi.org/10.1016/j.physletb.2007.10.057}.

\bibitem[{\citenamefont{Aguilar-Saavedra}(2009{\natexlab{a}})}]{AguilarSaavedra:2008zc}
\bibinfo{author}{\bibfnamefont{J.~A.} \bibnamefont{Aguilar-Saavedra}},
  \bibinfo{journal}{Nucl. Phys. B} \textbf{\bibinfo{volume}{812}},
  \bibinfo{pages}{181} (\bibinfo{year}{2009}{\natexlab{a}}),
  \eprint{0811.3842}.

\bibitem[{\citenamefont{Aguilar-Saavedra}(2009{\natexlab{b}})}]{AguilarSaavedra:2009mx}
\bibinfo{author}{\bibfnamefont{J.~A.} \bibnamefont{Aguilar-Saavedra}},
  \bibinfo{journal}{Nucl. Phys. B} \textbf{\bibinfo{volume}{821}},
  \bibinfo{pages}{215} (\bibinfo{year}{2009}{\natexlab{b}}),
  \eprint{0904.2387}.

\bibitem[{\citenamefont{Zhang and Willenbrock}(2011)}]{Zhang:2010dr}
\bibinfo{author}{\bibfnamefont{C.}~\bibnamefont{Zhang}} \bibnamefont{and}
  \bibinfo{author}{\bibfnamefont{S.}~\bibnamefont{Willenbrock}},
  \bibinfo{journal}{Phys. Rev. D} \textbf{\bibinfo{volume}{83}},
  \bibinfo{pages}{034006} (\bibinfo{year}{2011}), \eprint{1008.3869}.

\bibitem[{\citenamefont{Degrande et~al.}(2018)\citenamefont{Degrande, Maltoni,
  Mimasu, Vryonidou, and Zhang}}]{Degrande_2018}
\bibinfo{author}{\bibfnamefont{C.}~\bibnamefont{Degrande}},
  \bibinfo{author}{\bibfnamefont{F.}~\bibnamefont{Maltoni}},
  \bibinfo{author}{\bibfnamefont{K.}~\bibnamefont{Mimasu}},
  \bibinfo{author}{\bibfnamefont{E.}~\bibnamefont{Vryonidou}},
  \bibnamefont{and} \bibinfo{author}{\bibfnamefont{C.}~\bibnamefont{Zhang}},
  \bibinfo{journal}{Journal of High Energy Physics}
  \textbf{\bibinfo{volume}{2018}} (\bibinfo{year}{2018}), ISSN
  \bibinfo{issn}{1029-8479},
  \urlprefix\url{http://dx.doi.org/10.1007/JHEP10(2018)005}.

\bibitem[{\citenamefont{Cao et~al.}(2017)\citenamefont{Cao, Yan, Yu, and
  Zhang}}]{Cao:2015doa}
\bibinfo{author}{\bibfnamefont{Q.-H.} \bibnamefont{Cao}},
  \bibinfo{author}{\bibfnamefont{B.}~\bibnamefont{Yan}},
  \bibinfo{author}{\bibfnamefont{J.-H.} \bibnamefont{Yu}}, \bibnamefont{and}
  \bibinfo{author}{\bibfnamefont{C.}~\bibnamefont{Zhang}},
  \bibinfo{journal}{Chin. Phys. C} \textbf{\bibinfo{volume}{41}},
  \bibinfo{pages}{063101} (\bibinfo{year}{2017}), \eprint{1504.03785}.

\bibitem[{\citenamefont{Xie and Yan}(2021)}]{Xie:2021xtl}
\bibinfo{author}{\bibfnamefont{K.-P.} \bibnamefont{Xie}} \bibnamefont{and}
  \bibinfo{author}{\bibfnamefont{B.}~\bibnamefont{Yan}} (\bibinfo{year}{2021}),
  \eprint{2104.12689}.

\bibitem[{\citenamefont{Brivio and Trott}(2019)}]{Brivio_2019}
\bibinfo{author}{\bibfnamefont{I.}~\bibnamefont{Brivio}} \bibnamefont{and}
  \bibinfo{author}{\bibfnamefont{M.}~\bibnamefont{Trott}},
  \bibinfo{journal}{Physics Reports} \textbf{\bibinfo{volume}{793}},
  \bibinfo{pages}{1–98} (\bibinfo{year}{2019}), ISSN
  \bibinfo{issn}{0370-1573},
  \urlprefix\url{http://dx.doi.org/10.1016/j.physrep.2018.11.002}.

\bibitem[{\citenamefont{Brivio et~al.}(2017)\citenamefont{Brivio, Jiang, and
  Trott}}]{Brivio:2017btx}
\bibinfo{author}{\bibfnamefont{I.}~\bibnamefont{Brivio}},
  \bibinfo{author}{\bibfnamefont{Y.}~\bibnamefont{Jiang}}, \bibnamefont{and}
  \bibinfo{author}{\bibfnamefont{M.}~\bibnamefont{Trott}},
  \bibinfo{journal}{JHEP} \textbf{\bibinfo{volume}{12}}, \bibinfo{pages}{070}
  (\bibinfo{year}{2017}), \eprint{1709.06492}.

\bibitem[{\citenamefont{Dedes et~al.}(2017)\citenamefont{Dedes, Materkowska,
  Paraskevas, Rosiek, and Suxho}}]{SMEFTFR:2017zog}
\bibinfo{author}{\bibfnamefont{A.}~\bibnamefont{Dedes}},
  \bibinfo{author}{\bibfnamefont{W.}~\bibnamefont{Materkowska}},
  \bibinfo{author}{\bibfnamefont{M.}~\bibnamefont{Paraskevas}},
  \bibinfo{author}{\bibfnamefont{J.}~\bibnamefont{Rosiek}}, \bibnamefont{and}
  \bibinfo{author}{\bibfnamefont{K.}~\bibnamefont{Suxho}},
  \bibinfo{journal}{JHEP} \textbf{\bibinfo{volume}{06}}, \bibinfo{pages}{143}
  (\bibinfo{year}{2017}), \eprint{1704.03888}.

\bibitem[{\citenamefont{Buchmuller and Wyler}(1986)}]{Buchmuller:1985jz}
\bibinfo{author}{\bibfnamefont{W.}~\bibnamefont{Buchmuller}} \bibnamefont{and}
  \bibinfo{author}{\bibfnamefont{D.}~\bibnamefont{Wyler}},
  \bibinfo{journal}{Nucl. Phys. B} \textbf{\bibinfo{volume}{268}},
  \bibinfo{pages}{621} (\bibinfo{year}{1986}).

\bibitem[{\citenamefont{Gao and Berger}(2020)}]{Gao:2020ejr}
\bibinfo{author}{\bibfnamefont{J.}~\bibnamefont{Gao}} \bibnamefont{and}
  \bibinfo{author}{\bibfnamefont{E.~L.} \bibnamefont{Berger}}
  (\bibinfo{year}{2020}), \eprint{2005.12936}.

\bibitem[{\citenamefont{Alwall et~al.}(2014)\citenamefont{Alwall, Frederix,
  Frixione, Hirschi, Maltoni, Mattelaer, Shao, Stelzer, Torrielli, and
  Zaro}}]{Alwall:2014hca}
\bibinfo{author}{\bibfnamefont{J.}~\bibnamefont{Alwall}},
  \bibinfo{author}{\bibfnamefont{R.}~\bibnamefont{Frederix}},
  \bibinfo{author}{\bibfnamefont{S.}~\bibnamefont{Frixione}},
  \bibinfo{author}{\bibfnamefont{V.}~\bibnamefont{Hirschi}},
  \bibinfo{author}{\bibfnamefont{F.}~\bibnamefont{Maltoni}},
  \bibinfo{author}{\bibfnamefont{O.}~\bibnamefont{Mattelaer}},
  \bibinfo{author}{\bibfnamefont{H.~S.} \bibnamefont{Shao}},
  \bibinfo{author}{\bibfnamefont{T.}~\bibnamefont{Stelzer}},
  \bibinfo{author}{\bibfnamefont{P.}~\bibnamefont{Torrielli}},
  \bibnamefont{and} \bibinfo{author}{\bibfnamefont{M.}~\bibnamefont{Zaro}},
  \bibinfo{journal}{JHEP} \textbf{\bibinfo{volume}{07}}, \bibinfo{pages}{079}
  (\bibinfo{year}{2014}), \eprint{1405.0301}.

\bibitem[{\citenamefont{Alloul et~al.}(2014)\citenamefont{Alloul, Christensen,
  Degrande, Duhr, and Fuks}}]{Alloul:2013bka}
\bibinfo{author}{\bibfnamefont{A.}~\bibnamefont{Alloul}},
  \bibinfo{author}{\bibfnamefont{N.~D.} \bibnamefont{Christensen}},
  \bibinfo{author}{\bibfnamefont{C.}~\bibnamefont{Degrande}},
  \bibinfo{author}{\bibfnamefont{C.}~\bibnamefont{Duhr}}, \bibnamefont{and}
  \bibinfo{author}{\bibfnamefont{B.}~\bibnamefont{Fuks}},
  \bibinfo{journal}{Comput. Phys. Commun.} \textbf{\bibinfo{volume}{185}},
  \bibinfo{pages}{2250} (\bibinfo{year}{2014}), \eprint{1310.1921}.

\bibitem[{\citenamefont{Dordevic}(2019)}]{Dordevic:2019fft}
\bibinfo{author}{\bibfnamefont{M.}~\bibnamefont{Dordevic}}
  (\bibinfo{collaboration}{CMS}), \bibinfo{journal}{EPJ Web Conf.}
  \textbf{\bibinfo{volume}{222}}, \bibinfo{pages}{01001}
  (\bibinfo{year}{2019}).

\bibitem[{\citenamefont{Dawson et~al.}(2013)}]{Dawson:2013bba}
\bibinfo{author}{\bibfnamefont{S.}~\bibnamefont{Dawson}} \bibnamefont{et~al.}
  (\bibinfo{year}{2013}), \eprint{1310.8361}.

\bibitem[{\citenamefont{Peskin and Takeuchi}(1992)}]{Peskin:1991sw}
\bibinfo{author}{\bibfnamefont{M.~E.} \bibnamefont{Peskin}} \bibnamefont{and}
  \bibinfo{author}{\bibfnamefont{T.}~\bibnamefont{Takeuchi}},
  \bibinfo{journal}{Phys. Rev. D} \textbf{\bibinfo{volume}{46}},
  \bibinfo{pages}{381} (\bibinfo{year}{1992}).

\bibitem[{\citenamefont{Han and Skiba}(2005)}]{Han:2004az}
\bibinfo{author}{\bibfnamefont{Z.}~\bibnamefont{Han}} \bibnamefont{and}
  \bibinfo{author}{\bibfnamefont{W.}~\bibnamefont{Skiba}},
  \bibinfo{journal}{Phys. Rev. D} \textbf{\bibinfo{volume}{71}},
  \bibinfo{pages}{075009} (\bibinfo{year}{2005}), \eprint{hep-ph/0412166}.

\bibitem[{\citenamefont{Tanabashi et~al.}(2018)}]{Tanabashi:2018oca}
\bibinfo{author}{\bibfnamefont{M.}~\bibnamefont{Tanabashi}}
  \bibnamefont{et~al.} (\bibinfo{collaboration}{Particle Data Group}),
  \bibinfo{journal}{Phys. Rev. D} \textbf{\bibinfo{volume}{98}},
  \bibinfo{pages}{030001} (\bibinfo{year}{2018}).

\bibitem[{\citenamefont{Kane et~al.}(1992)\citenamefont{Kane, Ladinsky, and
  Yuan}}]{Kane:1991bg}
\bibinfo{author}{\bibfnamefont{G.~L.} \bibnamefont{Kane}},
  \bibinfo{author}{\bibfnamefont{G.~A.} \bibnamefont{Ladinsky}},
  \bibnamefont{and} \bibinfo{author}{\bibfnamefont{C.~P.} \bibnamefont{Yuan}},
  \bibinfo{journal}{Phys. Rev. D} \textbf{\bibinfo{volume}{45}},
  \bibinfo{pages}{124} (\bibinfo{year}{1992}).

\bibitem[{\citenamefont{Mahlon and Parke}(1997)}]{Mahlon:1996pn}
\bibinfo{author}{\bibfnamefont{G.}~\bibnamefont{Mahlon}} \bibnamefont{and}
  \bibinfo{author}{\bibfnamefont{S.~J.} \bibnamefont{Parke}},
  \bibinfo{journal}{Phys. Rev. D} \textbf{\bibinfo{volume}{55}},
  \bibinfo{pages}{7249} (\bibinfo{year}{1997}), \eprint{hep-ph/9611367}.

\bibitem[{\citenamefont{Cao and Yuan}(2005)}]{Cao:2004ky}
\bibinfo{author}{\bibfnamefont{Q.-H.} \bibnamefont{Cao}} \bibnamefont{and}
  \bibinfo{author}{\bibfnamefont{C.~P.} \bibnamefont{Yuan}},
  \bibinfo{journal}{Phys. Rev. D} \textbf{\bibinfo{volume}{71}},
  \bibinfo{pages}{054022} (\bibinfo{year}{2005}), \eprint{hep-ph/0408180}.

\bibitem[{\citenamefont{Sullivan}(2004)}]{Sullivan:2004ie}
\bibinfo{author}{\bibfnamefont{Z.}~\bibnamefont{Sullivan}},
  \bibinfo{journal}{Phys. Rev. D} \textbf{\bibinfo{volume}{70}},
  \bibinfo{pages}{114012} (\bibinfo{year}{2004}), \eprint{hep-ph/0408049}.

\bibitem[{\citenamefont{Cao et~al.}(2005)\citenamefont{Cao, Schwienhorst,
  Benitez, Brock, and Yuan}}]{Cao:2005pq}
\bibinfo{author}{\bibfnamefont{Q.-H.} \bibnamefont{Cao}},
  \bibinfo{author}{\bibfnamefont{R.}~\bibnamefont{Schwienhorst}},
  \bibinfo{author}{\bibfnamefont{J.~A.} \bibnamefont{Benitez}},
  \bibinfo{author}{\bibfnamefont{R.}~\bibnamefont{Brock}}, \bibnamefont{and}
  \bibinfo{author}{\bibfnamefont{C.~P.} \bibnamefont{Yuan}},
  \bibinfo{journal}{Phys. Rev. D} \textbf{\bibinfo{volume}{72}},
  \bibinfo{pages}{094027} (\bibinfo{year}{2005}), \eprint{hep-ph/0504230}.

\bibitem[{\citenamefont{Sirunyan et~al.}(2020)}]{Sirunyan:2018rlu}
\bibinfo{author}{\bibfnamefont{A.~M.} \bibnamefont{Sirunyan}}
  \bibnamefont{et~al.} (\bibinfo{collaboration}{CMS}), \bibinfo{journal}{Phys.
  Lett. B} \textbf{\bibinfo{volume}{800}}, \bibinfo{pages}{135042}
  (\bibinfo{year}{2020}), \eprint{1812.10514}.

\bibitem[{\citenamefont{Aaboud et~al.}(2017)}]{Aaboud:2016ymp}
\bibinfo{author}{\bibfnamefont{M.}~\bibnamefont{Aaboud}} \bibnamefont{et~al.}
  (\bibinfo{collaboration}{ATLAS}), \bibinfo{journal}{JHEP}
  \textbf{\bibinfo{volume}{04}}, \bibinfo{pages}{086} (\bibinfo{year}{2017}),
  \eprint{1609.03920}.

\bibitem[{\citenamefont{{CMS Collaboration}}(2019)}]{CMS:2019dvz}
\bibinfo{author}{\bibnamefont{{CMS Collaboration}}}
  (\bibinfo{collaboration}{CMS}) (\bibinfo{year}{2019}),
  \eprint{CMS-PAS-TOP-17-023}.

\bibitem[{\citenamefont{Khachatryan et~al.}(2016)}]{Khachatryan:2015dzz}
\bibinfo{author}{\bibfnamefont{V.}~\bibnamefont{Khachatryan}}
  \bibnamefont{et~al.} (\bibinfo{collaboration}{CMS}), \bibinfo{journal}{JHEP}
  \textbf{\bibinfo{volume}{04}}, \bibinfo{pages}{073} (\bibinfo{year}{2016}),
  \eprint{1511.02138}.

\bibitem[{\citenamefont{Tiko}(2016)}]{Tiko:2016brs}
\bibinfo{author}{\bibfnamefont{A.}~\bibnamefont{Tiko}}
  (\bibinfo{collaboration}{CMS}) (\bibinfo{year}{2016}), \eprint{1611.09397}.

\bibitem[{\citenamefont{Nocera et~al.}(2020)\citenamefont{Nocera, Ubiali, and
  Voisey}}]{Nocera:2019wyk}
\bibinfo{author}{\bibfnamefont{E.~R.} \bibnamefont{Nocera}},
  \bibinfo{author}{\bibfnamefont{M.}~\bibnamefont{Ubiali}}, \bibnamefont{and}
  \bibinfo{author}{\bibfnamefont{C.}~\bibnamefont{Voisey}},
  \bibinfo{journal}{JHEP} \textbf{\bibinfo{volume}{05}}, \bibinfo{pages}{067}
  (\bibinfo{year}{2020}), \eprint{1912.09543}.

\bibitem[{\citenamefont{Sirunyan et~al.}(2019)}]{Sirunyan:2018lzm}
\bibinfo{author}{\bibfnamefont{A.~M.} \bibnamefont{Sirunyan}}
  \bibnamefont{et~al.} (\bibinfo{collaboration}{CMS}), \bibinfo{journal}{Phys.
  Rev. D} \textbf{\bibinfo{volume}{99}}, \bibinfo{pages}{092005}
  (\bibinfo{year}{2019}), \eprint{1811.09696}.

\bibitem[{\citenamefont{Aad et~al.}(2020)}]{Aad:2020ivc}
\bibinfo{author}{\bibfnamefont{G.}~\bibnamefont{Aad}} \bibnamefont{et~al.}
  (\bibinfo{collaboration}{ATLAS}), \bibinfo{journal}{Phys. Rev. Lett.}
  \textbf{\bibinfo{volume}{125}}, \bibinfo{pages}{061802}
  (\bibinfo{year}{2020}), \eprint{2004.04545}.

\end{thebibliography}

\end{document}